\documentclass[aps,twocolumn]{revtex4}
\usepackage{epsfig}
\usepackage{psfrag}
\newcommand{\be}{\begin{equation}}
\newcommand{\ee}{\end{equation}}
\newcommand{\ba}{\begin{eqnarray}}
\newcommand{\ea}{\end{eqnarray}}
\newcommand{\nn}{$\langle n \rangle$}
\newcommand{\Dn}{$\Delta n\;$}

\begin{document}

\title{Scaling of loop-erased walks in 2 to 4 dimensions}

\author{Peter Grassberger}

\affiliation{John-von-Neumann Institute for Computing, Forschungszentrum J\"ulich,
D-52425 J\"ulich, Germany\\ and \\ Department of Physics and Astrophysics, University 
of Calgary, Alberta, Canada T2N 1N4}

\date{\today}

\begin{abstract}
We simulate loop-erased random walks on simple (hyper-)cubic lattices of dimensions
2, 3 and 4. These simulations were mainly motivated to test recent two loop 
renormalization group predictions for logarithmic corrections in $d=4$, simulations 
in lower dimensions were done for completeness and in order to test the algorithm.
In $d=2$, we verify with high precision the prediction $D=5/4$, where the number of 
steps $n$ after erasure scales with the number $N$ of steps before erasure as 
$n\sim N^{D/2}$. In $d=3$ we again find a power law, but with an exponent different 
from the one found in the most precise previous simulations: $D = 1.6236\pm 0.0004$.
Finally, we see clear deviations from the naive scaling $n\sim N$ in $d=4$. While they 
agree only qualitatively with the leading logarithmic corrections predicted by several 
authors, their agreement with the two-loop prediction is nearly perfect.
\end{abstract}

\maketitle

\section{Introduction}

The loop-erased random walk (LERW) is one of the simplest critical phenomena. 
It has no direct physical realization, although it is related to several well 
studied problems in statistical physics \cite{Majumdar,Duplantier,Schramm}. It 
was first studied by Lawler \cite{Lawler} as a simplified version of self 
avoiding walks. It is defined by performing a standard random walk, and 
erasing any loop as soon as it is formed. Thus it has no self-intersections, 
but it has different statistics from the usual self avoiding walk (SAW) where
the entire walk is erased as soon as a loop is formed.

Since there is no attrition for the LERW, the entropic critical exponent
(called $\gamma$ for SAWs) is trivially equal to 1, and any scaling behavior 
refers to geometric quantities. Let us denote by $N$ the number of steps of 
the original walk (without erasure), $n$ the number of steps after erasure, 
and $R$ any characteristic length scale such as the end-to-end or the 
gyration radius. Obviously, 
\be 
    \langle R^2\rangle \sim N.
\ee
Non-trivial scaling relates $n$ to $N$ or to $R$,
\be
   \langle n \rangle \sim N^{D/2} \sim \langle R^D\rangle,
\ee
where $D$ is the fractal dimension, or 
\be 
    \langle R\rangle \sim n^\nu
\ee
with $\nu = 1/D$.

It is known that the upper critical dimension for the LERW is $d=4$, with 
$D=2$ for $d>4$. It is also known that the LERW is related to spanning 
trees, which gives $D = 5/4$ for $d=2$ \cite{Majumdar,Duplantier,Schramm}.
For $d=4$ there are logarithmic corrections, which to leading 
order were given exactly by Lawler \cite{Lawler2} as 
\be
   \langle n \rangle \sim N / (\ln N)^{1/3}.
                    \label{lawler}
\ee
For the next-to-leading (two loop) logarithmic corrections at $d=4$, a functional 
renormalization group (FRG) was proposed in \cite{Fedorenko} which gives
\be
   \langle n \rangle \sim 
      N (\ln N)^{-1/3} \left[1+{4\ln\ln N \over 9\ln N} + O({1\over \ln N})\right].
                     \label{twoloop}
\ee

Finally, for $d=3$ the FRG gives in two loop approximation $D=1.614\pm 0.012$ 
\cite{Fedorenko}, which is in good agreement with the best simulation result
\cite{Agrawal}
\be
   D=1.6183\pm 0.0004 \quad (d=3, \;\; {\rm Ref.~[7]}).
                     \label{agrawal}
\ee

Since Eq.(\ref{twoloop}) represents a decisive test of the FRG, we decided to 
verify it by means of Monte Carlo simulations. Once we do this, it was deemed 
appropriate to simulate LERWs also in $d=2$ and $d=3$, in order to test the 
algorithm and to verify Eq.(\ref{agrawal}).

\section{Simulations}
 
In agreement with previous authors~\cite{Bursill,Agrawal}, we found that 
most of the scaling laws (1) to (3) (and similar other ones) are not well 
suited for precise estimates of the critical exponents, with one exception. 
By far the best suited is the first part of Eq.~(2),
$\langle n\rangle \sim N^{D/2}$. In the following we shall only consider this 
relation. In contrast, any measurements involving spatial extent, such as 
$\langle R^2\rangle$ versus $n$ or $\langle n\rangle$ versus $R^2$, gave 
very large errors \cite{Bursill}.

In order to detect loops, we have to store somehow the complete information 
about the entire LERW. As noted in \cite{Agrawal}, it is imperative to use 
long walks, for which a representation in terms of a simple bit map is no 
longer feasible. In the latter, we would use a Boolean variable $s_i$ for each 
site $i$ of the lattice, and would put $s_i=0$ if site $i$ has not yet 
been visited, and $s_i=1$ otherwise. Using present day 
computers with $\approx 10$ GB of memory would then allow $4-d$ lattices of 
at most $\approx 500^4$ sites, which would then allow only walks of $< 10^5$ 
steps without encountering finite lattice size corrections. Instead, we 
shall present below results for $N$ up to $2^{28}$ (for $d=3$) and $2^{27}$
(for $d=4$). This is only possible by using hashing. 

We use a different hashing strategy from that used in \cite{Agrawal}. Our 
hashing method had been
used before by us for a number of other statistical physics problems in 
high dimensions \cite{SAW,percol,dirperc}. It uses a virtual 
lattice with $2^{64}$ sites and with helical boundary conditions. In this 
lattice, each site is encoded by a single 64-bit integer, and neighbors of 
site $i$ are sites $i\pm 1, i\pm L_1, \ldots i\pm L_{d-1}$, all modulo $2^{64}$. 
Here the constants $L_k$ are of order $2^{64k/d}$, but are odd and not close 
to multiples of $2^p$ with large $p$. The hash function is simply obtained by 
using the last $m$ bits of $i$, with $m$ chosen so that $n < 2^m$ for the 
longest walks to be simulated. Collisions are resolved by means of a linked list. 
For random numbers we used Ziff's four-tap shift register \cite{Ziff}.

In each dimension, the number of walks with maximal $N$ was between $3\times 10^4$
and $10^6$, while there were roughly $2\times 10^7$ or more shorter walks, with 
$N < 10^4$ (see Table 1). The total CPU time used for these simulations was 
about three months on fast work stations.

\begin{table*}
\begin{center}
\begin{tabular}{|r|ccc|ccc|ccc|} \hline
          &   \multicolumn{3}{c|}{$d=2$}    &   \multicolumn{3}{c|}{$d=3$}   &  \multicolumn{3}{c|}{$d=4$}         \\ \hline
      $N$ &     \nn    & \#(walks)&   \Dn     &    \nn     &\#(walks) &  \Dn     &     \nn     &\#(walks) &  \Dn          \\ \hline
        1 &    1.00000 & 21591006 &  0.0000 &    1.00000 & 56646433 &  0.0000  &      1.00000 & 23103789 & 0.0000  \\
        2 &    1.50019 & 21591006 &  0.5791 &    1.66634 & 56646433 &  0.4403  &      1.75010 & 23103789 & 0.3847  \\
        4 &    2.49939 & 21591006 &  0.5451 &    2.98131 & 56646433 &  0.4165  &      3.23438 & 23103789 & 0.3470  \\
        8 &    4.06995 & 21591006 &  0.5315 &    5.28765 & 56646433 &  0.3948  &      5.99299 & 23103789 & 0.3163  \\
       16 &    6.52103 & 21591006 &  0.5170 &    9.31951 & 56646433 &  0.3785  &     11.15062 & 23103789 & 0.2907  \\
       32 &   10.32850 & 21591006 &  0.5094 &   16.36508 & 56646433 &  0.3656  &     20.85414 & 23103789 & 0.2655  \\
       64 &   16.22481 & 21591006 &  0.5030 &   28.67976 & 56646433 &  0.3552  &     39.21221 & 23103789 & 0.2442  \\
      128 &   25.33179 & 21591006 &  0.4984 &   50.21404 & 56646433 &  0.3468  &     74.13263 & 23103789 & 0.2261  \\
      256 &   39.39338 & 21591006 &  0.4953 &   87.88161 & 56646433 &  0.3410  &    140.89184 & 23103789 & 0.2107  \\
      512 &   61.09356 & 21591006 &  0.4936 &  153.81925 & 56646433 &  0.3368  &    269.01807 & 23103789 & 0.1972  \\
     1024 &   94.56645 & 21591006 &  0.4925 &  269.34263 & 56646433 &  0.3335  &    515.81337 & 23103789 & 0.1856  \\
     2048 &  146.20084 & 21591006 &  0.4911 &  471.83875 & 50426470 &  0.3311  &    992.71961 & 23103789 & 0.1756  \\
     4096 &  225.81969 & 21591006 &  0.4907 &  826.87931 & 50426470 &  0.3293  &   1916.83874 & 23103789 & 0.1669  \\
     8192 &  348.61386 & 21591006 &  0.4900 & 1449.44950 & 46915465 &  0.3281  &   3711.48360 & 23103789 & 0.1595  \\
    16384 &  538.04555 & 21591006 &  0.4902 & 2541.60437 & 46915465 &  0.3273  &   7204.79764 & 23103789 & 0.1526  \\
    32768 &  830.05473 & 21591006 &  0.4900 & 4458.15963 & 41260954 &  0.3265  &  14014.76907 & 18308386 & 0.1465  \\
    65536 &  1280.5943 & 21591006 &  0.4896 & 7820.39533 & 33299452 &  0.3259  &  27313.92469 & 18308386 & 0.1410  \\
   131072 &  1975.4362 & 15928658 &  0.4898 & 13720.6709 & 28158988 &  0.3257  &  53321.55335 & 12088423 & 0.1363  \\
   262144 &  3047.3501 & 15928658 &  0.4893 & 24080.9160 & 20197486 &  0.3248  &  104245.2345 &  6947959 & 0.1317  \\
   524288 &  4699.1856 & 11680341 &  0.4896 & 42264.1295 & 16335380 &  0.3257  &  204061.8368 &  4354748 & 0.1278  \\
  1048576 &  7246.2634 & 11680341 &  0.4893 & 74190.5005 & 10976331 &  0.3249  &  399949.6941 &  2621431 & 0.1239  \\
  2097152 &  11175.750 &  7933447 &  0.4893 & 130222.742 &  8796275 &  0.3247  &  784837.8597 &  1857338 & 0.1202  \\
  4194304 &  17238.505 &  5659211 &  0.4893 & 228620.387 &  3808255 &  0.3245  &  1541310.375 &  1146351 & 0.1167  \\
  8388608 &  26587.622 &  2757555 &  0.4894 & 401311.018 &  1883458 &  0.3247  &  3028942.200 &   715042 & 0.1165  \\
 16777216 &  41006.515 &  1340970 &  0.4899 & 704339.487 &  1441641 &  0.3253  &  5956568.655 &   290183 & 0.1120  \\
 33554432 &      --    &     --   &    --   & 1236822.13 &   977557 &  0.3247  &  11729958.86 &   165251 & 0.1098  \\
 67108864 &      --    &     --   &    --   & 2169238.10 &   513030 &  0.3246  &  23100632.37 &    89409 & 0.1073  \\
134217728 &      --    &     --   &    --   & 3808926.56 &   196378 &  0.3250  &  45529597.92 &    34590 & 0.1052  \\
268435456 &      --    &     --   &    --   & 6685906.62 &   134477 &  0.3244  &       --     &     --   &    --     \\ \hline
\end{tabular}
\caption{Number of walk realizations, average lengths, and standard deviations of $n$ for LERW in 2 to 4
   dimensions. $N$ is the number of steps including all loops, \Dn is the relative standard 
   deviation of $n$. Thus the standard deviation of the estimate of $\langle n \rangle$ is 
   $\langle n\rangle\Delta n/\sqrt{\rm \#(walks)}$).}
\end{center}
\end{table*}

\begin{figure}
\begin{center}
\psfig{file=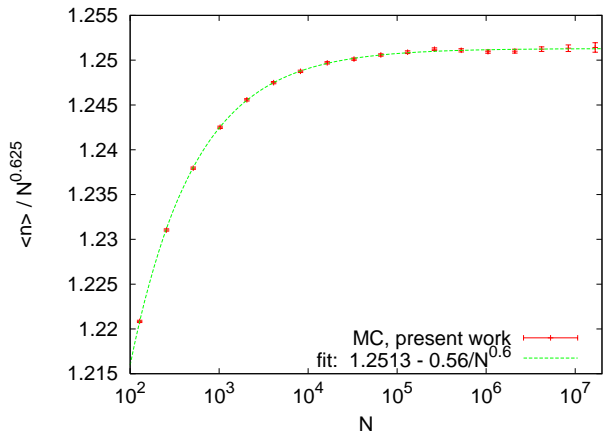,width=5.9cm, angle=270}
\caption{(color online) Plot of $\langle n\rangle / N^{5/8}$ against $\log N$, for $d=2$. 
   The smooth curve corresponds to a single correction term with exponent $\Delta = 0.6$. }
\end{center}
\end{figure}

\begin{figure}
\begin{center}
\psfig{file=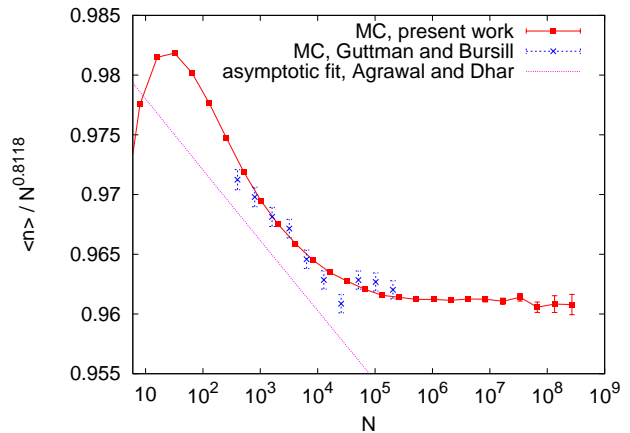,width=5.9cm, angle=270}
\caption{(color online) Plot of $\langle n\rangle / N^{0.8118}$ against $\log N$, for $d=3$. The 
   (nearly) straight line corresponds to the asymptotic behavior obtained with the estimate of $D$
   published in \cite{Agrawal}. The points with large error bars are the Monte Carlo simulations 
   of \cite{Bursill}.}
\end{center}
\end{figure}

\begin{figure}
\begin{center}
\psfig{file=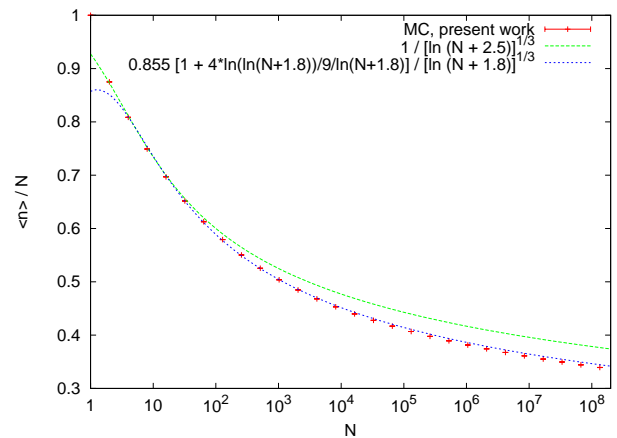,width=5.9cm, angle=270}
\caption{(color online) Plot of $\langle n\rangle / N$ against $\log N$, for $d=4$. The 
   two smooth curves are the leading log prediction and the two loop improvement. Both 
   contain unknown constants which are determined by fitting to the 
   data for $4\leq N \leq 16$.}
\end{center}
\end{figure}

\section{Results}

Our results are summarized in Table 1 and in Figs.~1 to 3. Figure 1 shows that our 
data for $d=2$ are perfectly consistent with $D=5/4$, as expected. It also shows 
that the corrections to scaling are quantitatively described by a single power with 
exponent $\Delta = 0.6$. There is to our knowledge no theoretical prediction for 
the latter, although it should be possible to obtain one from conformal invariance.

The data for $d=3$ shown in Fig.~2 are much more interesting. The first surprise 
is that a single power would not be enough to describe the corrections to scaling,
due to the lack of convexity of the curve $\langle n\rangle$ versus $\log N$. In 
view of this we refrain from quoting a value for the leading correction exponent, 
and we quote rather large errors for the scaling exponent:
\be
   D = 1.6236 \pm 0.0004 \quad (d=3).
\ee
Although this error is equal to the error quoted in Ref.~\cite{Agrawal}, we believe 
that we can firmly exclude the latter estimate, which is about 13 standard deviations 
away from our estimate. In order to illustrate the contradiction between our two 
estimates, we plotted in Fig.~2 also the asymptotic behavior based on the estimate of 
Ref.~\cite{Agrawal}. 

This estimate was obtained not by measuring $\langle n \rangle$
versus $N$, but by measuring the loop size distribution. To obtain $D$ from this, 
the authors of \cite{Agrawal} had to make a parameterized scaling ansatz for it. This 
ansatz had also to take into account that the loop size distribution has a cut off for 
finite $N$. 
It might be that either the ansatz was not general enough to take into account the 
finite-size corrections seen in Fig.~2, or that the cut off was parameterized wrongly. 
In contrast, the authors of Ref.~\cite{Bursill} used $\langle n\rangle$ versus $N$, as 
we do. They also cited their raw data, and as seen 
from Fig.~2 they are in perfect agreement with our simulations.

Finally, our data for $d=4$ are shown in Fig.~3. Without logarithmic corrections we would 
have $\langle n \rangle /N = const$. Equations~(\ref{lawler}) and (\ref{twoloop}) contain 
in principle also arbitrary integration constants and, in addition, the results of higher 
order corrections. This can be taken into account by replacing $N$ in a numerical 
analysis by $N/N_0$ or, alternatively, by $N+N_0$. A priori neither seems preferred. 
For both choices the constant $N_0$ is unknown and can take different values in 
Eqs.~(\ref{lawler}) and (\ref{twoloop}). We found that using the second (additive) 
choice gave better fits, and will use it in the following. We determined $N_0$ somewhat 
arbitrarily such that the MC data fitted the analytic expressions for $4\leq N \leq 16$.
In both cases (leading log and two loops) this gave $N_0$ roughly of order 1 (more 
precisely, 2.5 and 1.8). The main conclusion from Fig.~3 is that our MC data agree 
qualitatively with the leading log predictions, but not perfectly. This difference is 
nearly completely eliminated when the two-loop correction is included. There remains
a small residual difference, but we can definitely say that including the two-loop 
correction gives a big improvement and suggests that the FRG is basically correct.

\section{Conclusion}

Our simulations indicate clearly that the FRG analysis of Ref.~\cite{Fedorenko} is 
basically correct in four dimensions, where its predictions should be most reliable.
It is less successful in $d=3$, where it gives a too big change over the one-loop
result. The latter conclusion depends on our new estimate for the fractal LERW 
dimension in $d=3$, which is about 13 (new and old) standard deviations away from the 
best previous estimate. Finally, we verify the known value of the fractal dimension 
in $d=2$ with higher precision than previous Monte Carlo analyses, and we present 
estimates for the correction to scaling exponent in $d=2$. It should be possible
to calculate the latter analytically from conformal invariance.

Acknowledgments: I am indebted to Prof. Andrei Fedorenko for bringing this problem to 
my attention. I also want to thank him and Profs. Pierre Le Doussal, Kay Wiese and 
Deepak Dhar for very helpful correspondences.

\end{document}